# Design of Thin-Film-Transistor (TFT) arrays using current mirror circuits for Flat Panel Detectors (FPDs)

**Nur** SULTAN SALAHUDDIN [1,2], **Michel** PAINDAVOINE [1], **Nurul** HUDA [2*], **Michel** PARMENTIER [3]

[1] LE2I Laboratory UMR CNRS 5158, University of Burgundy, Dijon, France.
[2] Gunadarma University, Jakarta, Indonesia.
[3] Laboratoire Imagerie et Ingénierie pour la santé, University of Franche-Comte, Besançon, France.

*Abstract* -We designed 4x4 matrix TFTs arrays using current mirror amplifiers. Advantages of current mirror amplifiers are they need less requiring switches and the conversion time is short. The TFTs arrays 4x4 matrix using current mirror circuits have been fabricated and tested with success. The TFTs array directly can process signals coming from 16 pixels in the same node. This enables us to make the summation of the light intensities of close pixels during a reading.

## 1. Introduction

Flat panel detectors (FPDs) may be used in gamma cameras dedicated for medical applications. Flat panel imagers are based on solid-state integrated circuits (IC) technology, similar in many ways to the imaging chips used in visible wavelength digital photography and video. The heart of the flat panel digital detector consists of a two-dimensional array of amorphous silicon photodiodes and thin-film transistors (TFTs) [1], all deposited on a single substrate. Utilizing thin film technology similar to that used in the fabrication of integrated circuits, layers of amorphous silicon and various metals and insulators are deposited on a glass substrate to form the photodiodes and TFTs matrix, as well as the interconnections, and the contacts on the edges of the panel. The construction of FPDs is similar in many ways to flat panel displays, and uses many of the same technologies. Figure 1 shows the construction of a typical FPD [2]. At the core is an amorphous-silicon TFT/photodiode array.

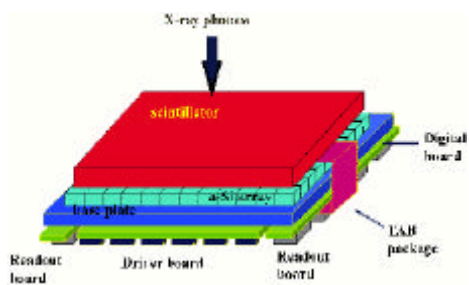

Figure 1. Flat panel detector internal architecture.

The TFT/photodiode matrix [2][3] is normally scanned progressively as shown in figure 2, one line at a time from top to bottom. The TFTs are essentially switches. When a large positive voltage is applied to one of the gate lines, the switches (TFTs) in the selected row are closed, causing them to conduct electricity. With the TFTs energized, each pixel in the selected row discharges the stored signal electrons onto the data line. At the end of each dataline is a charge integrating amplifier which converts the charge packet to a voltage. At this point the electronics systems vary by manufacturer, but generally there is a programmable gain stage and an Analog-to-Digital Converter (ADC), which converts the voltage to a digital number. One important aspect of the scanning is that the FPD is continuously collecting the entire incident signal; none is lost even during the discharge of the pixel. The FPD is an integrating detector and the integration time for each pixel is equal to the frame time.

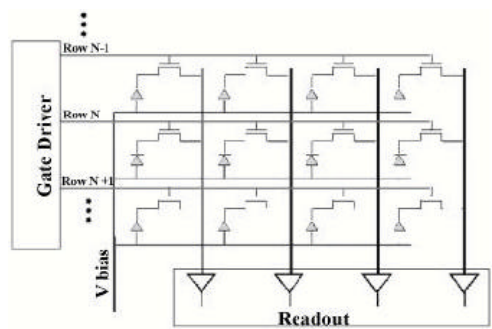

Figure 2. TFT/Photodiode array schematic.

We have designed 4x4 matrix TFTs using current mirror [4] amplifiers as shown in figure 3. Current mode was choosen because it's ussually offers the advantages of less requiring the switches (TFTs) and the conversion time is short. These characteristics will be particularly beneficial at the time of the design of the subsequent circuits on the way of data, mainly the converters analog/digital.

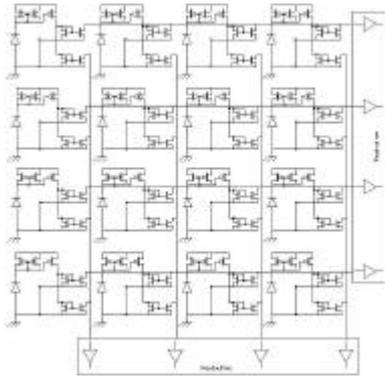

Figure 3  The 4x4 matix TFTs with current mirror.

But, more important still in our case, this mode enables us to calculate easily and instantaneously the average of the light intensities of neighbouring pixels in order to modify the resolution of the image. Indeed, we can sum the signals directly coming from several pixels in the same node. This enables us to make the summation of the light intensities of close pixels during a reading.

## 2. Pixel Design Description

The circuit of active matrix composed of one line and one row with active pixel is illustrated figure 4. It is a completely symmetrical circuit, established in each one pixel of the matrix. It includes a PMOS current mirror (M1-M3 are located in pixel), two NMOS current mirror (M4-M5 and M4' -M5 ') are common for one line and one row M6-M10 and M6' -M10 ' are the circuits of the amplifier.

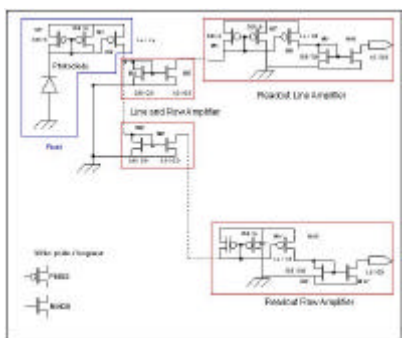

Figure 4 The Circuit of active marix.

## 3. Fabrication and Test Results

The circuits have been designed using the Mentor Graphics CAD tools [5] in combination with the AMS CMOS 0.6 µm design kit. This technology uses one layer of ploysilicium and three layers of metal. Figure 5 shows the drawing of this circuit, realized with the Mentor software. Figure 6 represents Photograph of a TFTs chip.

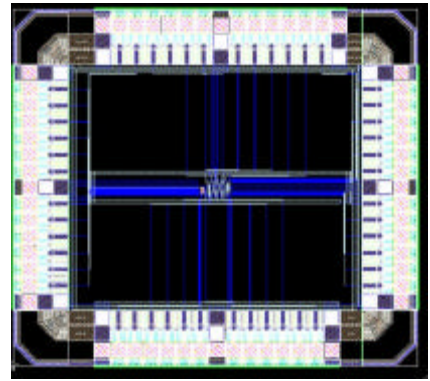

Figure 5  Layout of the TFTs with 4x4 matrix in 0.6 µm CMOS proses

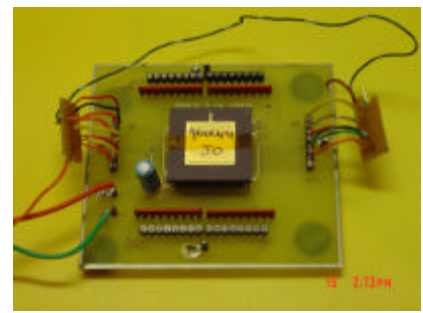

Figure 6  Photograph of the TFTs with 4x4 matrix in 0.6 µm CMOS proses

In order to the test  the circuits of active matrix of flat panel, we used several photodiodes where each of one is connected to one input of the circuit. Output of the circuit are connected to a matrix of LEDs in order to validate this circuit. Figure 7 illustrates the validation method.

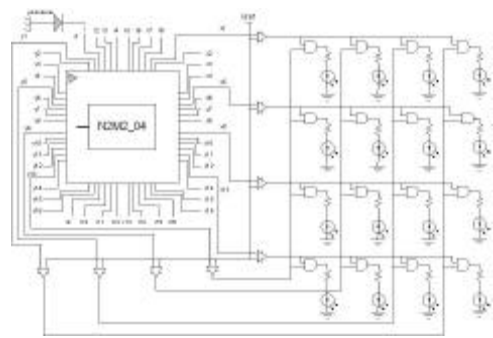

Figure 7  Experimental  of  active matix

The principle of the circuit is that if  value of the pixel is higher than the value of reference of tension (Vr), the exit of the comparator is in a high state (5V). On the other hand, if value of the pixel is less than the value of the tension Vr, the exit of the comparator is in a low state (0V). After the comparison, the exits of the comparator are entered to the logical gating circuit AND. In the logical gating circuit AND, if signal comparator from line and row are 5V and corresponding LED is active.

After the test of the response of our active matrix, we have analysed the dynamic response of active matrix. We used the luminescent blue diode (source) driven with a impulse generator. Figure.8, shows the results of measurement of the dynamic responses of active matrix with a blue LED calibrated for an illumination of 0.4 lux

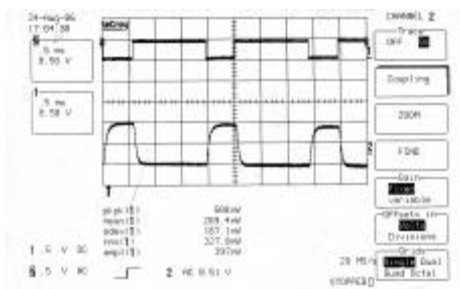

Figure 8  Response in dynamic mode  of active matix

We measured the signals directly coming from several pixels in the same node. Figure 9, shows the results of measurement of the dynamic responses of active matrix.

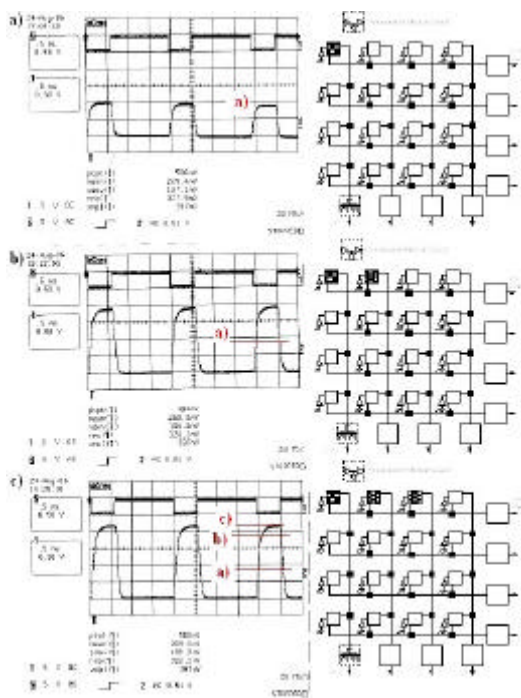

Figure 9  Response in dynamic response  of active matrix with several pixels

**Conclusions and Perspectives**

We have designed a 4x4 matrix TFTs using current mirror amplifiers for image sensor. We can sum directly the signals coming from several pixels in the same node. This enables us to make the summation of the light intensities of close pixels during a reading.

**References**

[1] GEs Digital Flat Panel Tecnology," Digital Flat Panel". (http://www.e-radiography.net/radtech/f/flat_panel.htm)
[2] VARIAN Medical System ,"Flat Panel X-Ray Imaging". (http://www.varian.com/xray/pdf/Flat%20Panel%20Xray%20Imaging%2011-11-04.pdf )
[3] Wei Zhao and J.A. Rowlands, ," Digital radiology using active matrix readout of amorphous selenium: Theoretical analysis of detective quantum efficiency," Medical Physics, Vol.24, No.12, December 1997.
[4] B. M. Wilamowski, E. S. Ferre-Pikal, O. Kaynak," Low power, current mode CMOS circuits for synthesis of arbitrary nonlinear fuctions," 9[th] NASA Symposium on VLSI Design 2000.
[5] Mentor Graphics Corporation. (http://www.mentor.com)